\documentclass[12pt]{emulateapj}
\usepackage{natbib}
\usepackage[normalem]{ulem} 

\newcommand{\etal}{et~al.}
\newcommand{\eg}{e.g.,}

\newcommand{\msun}{$M_\odot$}

\newcommand{\spitzer}{{\it Spitzer}}
\newcommand{\chandra}{{\it Chandra}}
\newcommand{\hst}{{\it HST}}

\newcommand{\cl}{IDCS~J1426.5+3508}
\newcommand{\clzeimann}{IDCS~J1433.2+3306}

\newcommand{\fgas}{$f_{gas}$}

\newcommand{\Mfive}{$M_{\mbox{\scriptsize $500$}}$} 
\newcommand{\Mfiveyx}{$M_{\mbox{\scriptsize $500,Y_X$}}$}
\newcommand{\Mfivetx}{$M_{\mbox{\scriptsize $500,T_X$}}$}
\newcommand{\Mfivelx}{$M_{\mbox{\scriptsize $500,L_X$}}$}
\newcommand{\Mfivemg}{$M_{\mbox{\scriptsize $500,M_g$}}$}
\newcommand{\Mfivesz}{$M_{\mbox{\scriptsize $500,SZ$}}$}

\newcommand{\Mfivearc}{$M_{\mbox{\scriptsize $500,{\rm arc}$}}$}
\newcommand{\Mfivelxbol}{$M_{\mbox{\scriptsize $500,L_{X,\rm bol}$}}$}

\newcommand{\rfive}{$r_{\mbox{\scriptsize 500}}$}

\newcommand{\rfivetx}{$r_{\mbox{\scriptsize $500,T_x$}}$}
\newcommand{\rfivemg}{$r_{\mbox{\scriptsize $500,M_g$}}$}

\newcommand{\Lxbol}{$L_{\mbox{\scriptsize $X,{\rm bol}$}}$}

\def\spose#1{\hbox to 0pt{#1\hss}}
\def\simlt{\mathrel{\spose{\lower 3pt\hbox{$\mathchar"218$}}
     \raise 2.0pt\hbox{$\mathchar"13C$}}}
\def\simgt{\mathrel{\spose{\lower 3pt\hbox{$\mathchar"218$}}
     \raise 2.0pt\hbox{$\mathchar"13E$}}}

\shorttitle{\cl: The Most Massive Galaxy Cluster at $z>1.5$}
\shortauthors{Brodwin et al.}

\begin{document}


\title{\cl: The Most Massive Galaxy Cluster at $z>1.5$}


\author{Mark~Brodwin\altaffilmark{1},
 Michael~McDonald\altaffilmark{2},
 Anthony~H.~Gonzalez\altaffilmark{3},
 S.~A.~Stanford\altaffilmark{4,5},
 Peter~R.~Eisenhardt\altaffilmark{6}, 
 Daniel~Stern\altaffilmark{6}
 \& Gregory~R.~Zeimann\altaffilmark{7}
}


\altaffiltext{1}{Department of Physics and Astronomy, University of Missouri, Kansas City, MO 64110}
\altaffiltext{2}{Kavli Institute for Astrophysics and Space Research, Massachusetts Institute of Technology, Cambridge, MA 02139}
\altaffiltext{3}{Department of Astronomy, University of Florida, Gainesville, FL 32611}
\altaffiltext{4}{Department of Physics, University of California, Davis, CA 95616}
\altaffiltext{5}{Institute of Geophysics and Planetary Physics, Lawrence Livermore National Laboratory, Livermore, CA 94551}
\altaffiltext{6}{Jet Propulsion Laboratory, California Institute of Technology, Pasadena, CA 91109}
\altaffiltext{7}{Department of Astronomy and Astrophysics, Pennsylvania State University, University Park, Pennsylvania 16802}



\begin{abstract}
  We present a deep (100\,ks) \chandra\ observation of \cl, a
  spectroscopically confirmed, infrared-selected galaxy cluster at $z
  =1.75$.  This cluster is the most massive galaxy cluster currently
  known at $z > 1.5$, based on existing Sunyaev-Zel’dovich (SZ) and
  gravitational lensing detections.
  We confirm this high mass via a variety of X-ray scaling relations,
  including T$_X$--M, $f_g$--M, Y$_X$--M, and L$_X$--M, finding a
  tight distribution of masses from these different methods, spanning
  M$_{500}$~=~2.3--3.3 $\times$ 10$^{14}$ M$_{\odot}$, with the
  low-scatter $Y_X$-based mass \Mfiveyx\ $= 2.6^{+1.5}_{-0.5} \times
  10^{14}$~M$_\odot$.
  \cl\ is currently the only cluster at $z > 1.5$ for which X-ray, SZ
  and gravitational lensing mass estimates exist, and these are in
  remarkably good agreement.
  We find a relatively tight distribution of the gas-to-total mass
  ratio, employing total masses from all of the aforementioned
  indicators, with values ranging from $f_{gas,500}$~=~0.087--0.12.
  We do not detect metals in the intracluster medium (ICM) of this
  system, placing a 2$\sigma$ upper limit of $Z(r<R_{500}) <
  0.18\,Z_{\odot}$.  This upper limit on the metallicity suggests that
  this system may still be in the process of enriching its ICM.
  The cluster has a dense, low-entropy core, offset by $\sim$30 kpc from
  the X-ray centroid, which makes it one of the few ``cool core''
  clusters discovered at $z>1$, and the first known cool core cluster
  at $z>1.2$. The offset of this core from the large-scale centroid
  suggests that this cluster has had a relatively recent
  ($\lesssim$\,500 Myr) merger/interaction with another massive
  system.
\end{abstract}





\section{Introduction}

In recent years the study of galaxy clusters has meaningfully entered
the $z>1$ regime, with several surveys identifying large samples via
X-ray \citep{fassbender11, mehrtens12}, Sunyaev-Zel'dovich
\citep[SZ,][]{bleem15, hasselfield13, planckSZ15}, infrared
\citep[IR,][]{eisenhardt08,muzzin09,papovich10, rettura14} and radio
(\citealt{wylezalek13,galametz13,blanton14}; R. Paterno-Mahler et
al. 2015, in preparation) selections.  These surveys have extended the
reach of cluster cosmology \citep[\eg][]{benson13}, scaling relations
\citep[\eg][]{andersson11} and galaxy formation and evolution in the
richest environments \citep[\eg][]{tran10, hilton10, brodwin13} to
$z\sim 1.5$.

It is crucial to identify the earliest massive progenitors of these $1
\la z \la 1.5$ cluster samples, and the present-day massive clusters
into which they evolve, in order to quantify the build-up of the
intracluster medium (ICM) and the establishment of self-similarity.
In particular, the scaling relations between different cluster mass
observables --- \eg\ ICM temperature, SZ signal, weak lensing --- are
calibrated to better than $\sim 20\%$ at $z < 0.5$
\citep[\eg][]{kravtsov06, vikhlinin09_cosmo, andersson11}, but are
currently poorly constrained at high redshifts ($z > 1$) where
clusters provide the greatest leverage as probes of the growth of
structure.

Very massive, high redshift galaxy clusters also provide a natural
testing ground to confirm or refute the predictions stemming from
recent galaxy evolution studies in clusters at $1 \la z \la 1.5$.
While little star formation activity is seen in the most massive ($M
\sim 10^{15}$ \msun) South Pole Telescope (SPT) clusters in this
regime \citep[\eg][]{brodwin10, foley11, stalder13}, the case is
notably different for clusters with masses in the range $M \sim (1-4)
\times 10^{14}$ \msun.  Indeed, \citet{brodwin13} reported the
discovery of a major epoch of star formation at $z\ga 1.4$ for IRAC
Shallow Cluster Survey \cite[ISCS,][]{eisenhardt08} clusters in this
mass range.  They attributed it in part to a high merging rate of
gas-rich cluster members, as suggested by \citet{mancone10}.  A
consequence of this model is that the epoch at which the merging (and
hence merger-induced star formation) ceases should be a function of
cluster mass.  While the SPT clusters at $z \la 1.3$ are generally too
massive to permit efficient merging, their lower-mass progenitors at
higher redshifts should have the high star formation rates seen in the
ISCS.  Indeed, in the most distant SPT cluster, SPT-CL J2040-4451,
with a mass of \Mfivesz\ $\sim 3.2 \times 10^{14}$ \msun\ at $z=1.48$,
residual star formation activity still persists after its earlier
bursting phase \citep{bayliss14}

To extend the infrared cluster search to $z>1.5$, we repeated the
methodology of the ISCS using the \spitzer\ Deep, Wide-Field Survey
\citep[SDWFS,][]{ashby09}, which quadrupled the \spitzer/IRAC exposure
time over the 9 deg$^2$ IRAC Shallow Survey \citep{eisenhardt04}.  The
resulting survey, the IRAC Distant Cluster Survey
\citep[IDCS,][]{stanford12, brodwin12, gonzalez12}, has identified two
of the most distant clusters to date: \cl\ at $z=1.75$
\citep{stanford12} and \clzeimann\ at $z=1.89$ \citep{zeimann12}.  The
latter appears to be a moderate mass cluster still in the process of
formation, whereas \cl\ at $z=1.75$, the subject of this paper, is a
very massive cluster.

\citet{stanford12} reported a detection in only 8.3 ks of archival
{\it Chandra X-ray Observatory} data, resulting in an $L_X$-based mass
estimate of \Mfivelx\ $= (3.3 \pm 1.0) \times 10^{14}$
\msun\footnote{$M_{\Delta,\rm A}$ refers to the mass within
  $r_\Delta$, the radius at which the mean overdensity is $\Delta$
  times the critical density, as inferred via mass proxy A.}.  \cl\
was also observed with the Sunyaev--Zel’dovich Array (SZA), a subarray
of the Combined Array for Research in Millimeter-wave Astronomy
(CARMA).  An SZ-based mass of \Mfivesz\ $ = (2.6 \pm 0.7) \times
10^{14}$ \msun\ was measured from the strong (5.3 $\sigma$) decrement
\citep{brodwin12}.  Finally, \cl\ has a giant gravitational arc, from
which a minimum mass of \Mfivearc\ $ \ga 2.0 \times 10^{14}$ \msun\
was estimated \citep{gonzalez12}.  Although these independent mass
measurements are all in good agreement, indicating that \cl\ is a very
massive, relaxed cluster at $z=1.75$, the $L_X$-based X-ray mass was
based on only 53 counts and is highly uncertain.

In this paper we present deep new \chandra\ observations from which we
measure the ICM properties of \cl.  In \textsection{\ref{Sec: Data}}
we describe the data used in this analysis.  In \textsection{\ref{Sec:
    Physical}} we present the flux, energy spectrum and gas density
profile of \cl, along with the quantities we derive from these direct
measurements, including the luminosity, temperature, gas mass and
metallicity.  Using standard scaling relations from the literature, we
estimate \Mfive\ for \cl\ from four different X-ray estimators in
\textsection{\ref{Sec: Masses}}, and compare these to complementary
SZ- and lensing-based mass estimates.  We also compute gas fractions
for each of these halo mass estimates and compare these with the value
predicted from low-redshift clusters.  In \textsection{\ref{Sec:
    Discussion}} we place \cl\ in the context of the known $z>1.5$
galaxy cluster population and discuss the evolutionary state of its
ICM.  Finally, we present our conclusions in \textsection{\ref{Sec:
    Conclusions}}.  We use $\Omega_m = 0.27$, $\Omega_\Lambda = 0.73$
and $H_0 = 70$ km s$^{-1}$ Mpc$^{-1}$ throughout.

\begin{figure*}[Hbthp]
\label{Fig: images}
\plotone{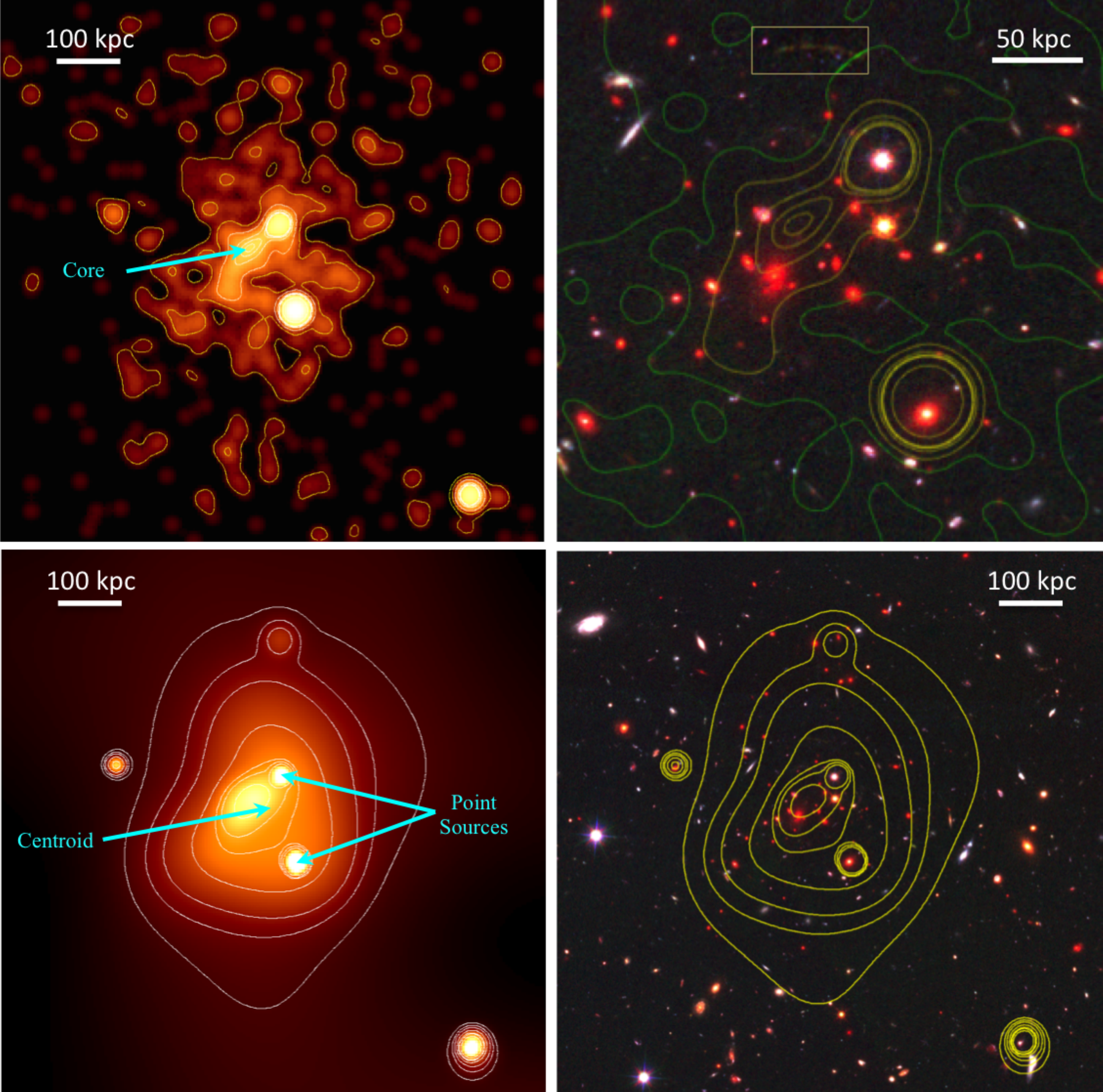}
\caption{{\it Upper left:} Gaussian-smoothed (FWHM = 3\arcsec)
  \chandra\ 0.5-2.0 keV image of \cl.  Contours correspond to 0.02,
  0.05, 0.1, 0.18, 0.22, 0.24, 0.25 and 0.45 counts per 0.492\arcsec\
  $\times$0.492\arcsec\ pixel.  The location of the core as traced by
  the ICM peak is indicated. {\it Upper right}: Pseudocolor ACS/F606W,
  ACS/F814W and WFC3/F160W \hst\ image of the central region of \cl,
  with the X-ray contours from the upper left panel overlaid.  The
  giant arc, boxed, is visible $\sim$125 kpc north of the BCG
  \citep[see][]{gonzalez12}.  {\it Lower left:} Adaptively smoothed
  image of \chandra\ 0.5-2.0 keV image of \cl, showing the large-scale
  structure from which the cluster centroid was measured.  The
  contours correspond to 0.011, 0.016, 0.022, 0.038, 0.065, 0.085 and
  0.15 counts per 0.492\arcsec\ $\times$ 0.492\arcsec\ pixel.  The
  peak emission in the core is offset by $\sim$30 kpc from the cluster
  centroid.  Two bright point sources discussed in the text are
  indicated.  {\it Lower right:} The \hst\ image with the same scale
  and contours as the adaptively smoothed X-ray image.}
\end{figure*}

\section{Data}
\label{Sec: Data}

The X-ray data used in this work were acquired via \chandra\ proposal
148000534 (PI: Brodwin). A total exposure of 100\,ks was acquired over
two pointings (OBSIDs 15168 and 16321).  This exposure time was chosen
to obtain 500 X-ray source counts, but due to the evolving effective
area at low energies \citep{marshall04, odell13} we obtained slightly
fewer than anticipated.  The data, obtained with ACIS-I, were cleaned
for background flares before applying the latest calibration
corrections using \textsc{ciao} v4.6 and \textsc{caldb} v4.6.1.1.

\section{X-ray Properties of \cl}
\label{Sec: Physical}

\subsection{Images, Centroid and Peak}

\chandra\ and \hst\ images of \cl\ are shown in Figure \ref{Fig:
  images}.  The first column shows \chandra\ images and contours for
Gaussian-smoothed (FWHM = 3\arcsec, {\it upper panel}) and adaptively
smoothed ({\it lower panel}) 0.5--2.0 keV X-ray images, respectively.
The 2nd column shows these contours overlaid on color optical/IR \hst\
images, a full description of which are given in (Mo \etal\ in prep.).
The upper right panel is zoomed in to the cluster core to better show
the brightest cluster galaxy (BCG) and the giant gravitational arc.

We use the centroid measured within a 250--500\ kpc annulus as the
cluster center, which tends to provide less biased estimates of the
global properties for unrelaxed clusters. This choice of center,
$(\alpha_X$, $\delta_X)$ = (14:26:32.6, +35:08:25), is within
4\arcsec\ of the X-ray peak, within 5\arcsec\ of the BCG position
\citep{stanford12} and within 28\arcsec\ of the SZ centroid
\citep{brodwin12}.  Given the uncertainty in the SZ centroid ($\approx
35\arcsec$), this positional offset is not statistically significant.

\subsection{Point Sources}

The incidence of AGN in clusters has been shown to increase rapidly
with redshift \citep{eastman07, galametz09, martini13, alberts15}.
This is natural consequence of the \citet{brodwin13} merger model
discussed above, where the mergers that induce starbusts also fuel
the AGN that provide the eventual quenching.

Point sources were identified and masked using an automated routine
following the wavelet decomposition technique described in
\citet{vikhlinin98}.  There are two bright X-ray point source near the
cluster core, previously discussed in \citet{stanford12}.  The
northern one is a QSO in the cluster, with a flux of $S_{0.5-2}
\approx 1.85_{-0.37}^{+0.33} \times 10^{-15}$ erg~s$^{-1}$~cm$^{-2}$.  The southern
one is a bright radio source listed in both the NVSS \citep{condon98}
and FIRST \citep{becker95} catalogs.  It has a flux of $\sim$95 mJy at
1.4 GHz, and was found to have a 31 GHz flux of $5.3 \pm 0.3$ mJy in
the SZ analysis of \citet{brodwin12}.  Here we measure a flux of
$S_{0.5-2} \approx 5.68_{-0.69}^{+0.71} \times 10^{-15}$ erg~s$^{-1}$~cm$^{-2}$.
These point sources, along with several others at larger
clustercentric radii, are masked and do not affect this analysis.

\subsection{Counts, Flux and Luminosity}
We measure 401 net counts (0.5--6\,keV) from \cl, after point source
masking and background subtraction. The flux in the soft band is
$S_{0.5-2} = (2.2 \pm 0.6) \times 10^{-14}$
erg~s$^{-1}$~cm$^{-2}$, corresponding to a luminosity of $L_{0.5-2} =
(3.6 \pm 0.5) \times 10^{44}$ erg~s$^{-1}$.  The bolometric
luminosity over the 0.01--100 keV energy range is measured from the
best-fit model described below (\textsection\ref{Sec: T}) to be
\Lxbol\ $= (12.8 \pm 1.1) \times 10^{44}$ erg~s$^{-1}$.  These
quantities are all measured within $r_{500,Y_X} = 530 \pm 10$ kpc, as
determined from the core-excised Y$_X$ measurement described below
(\textsection\ref{Sec: Yx}).  \citet{pratt09} showed the bolometric
luminosity to be a lower scatter mass proxy than $L_{0.5-2}$.  Using
their scaling relation, we find \Mfivelxbol\ = $(2.8 \pm 0.3) \times
10^{14}$ \msun.

\subsection{Temperature}
\label{Sec: T}
\begin{figure}[bthp]
\epsscale{1.2} 
\plotone{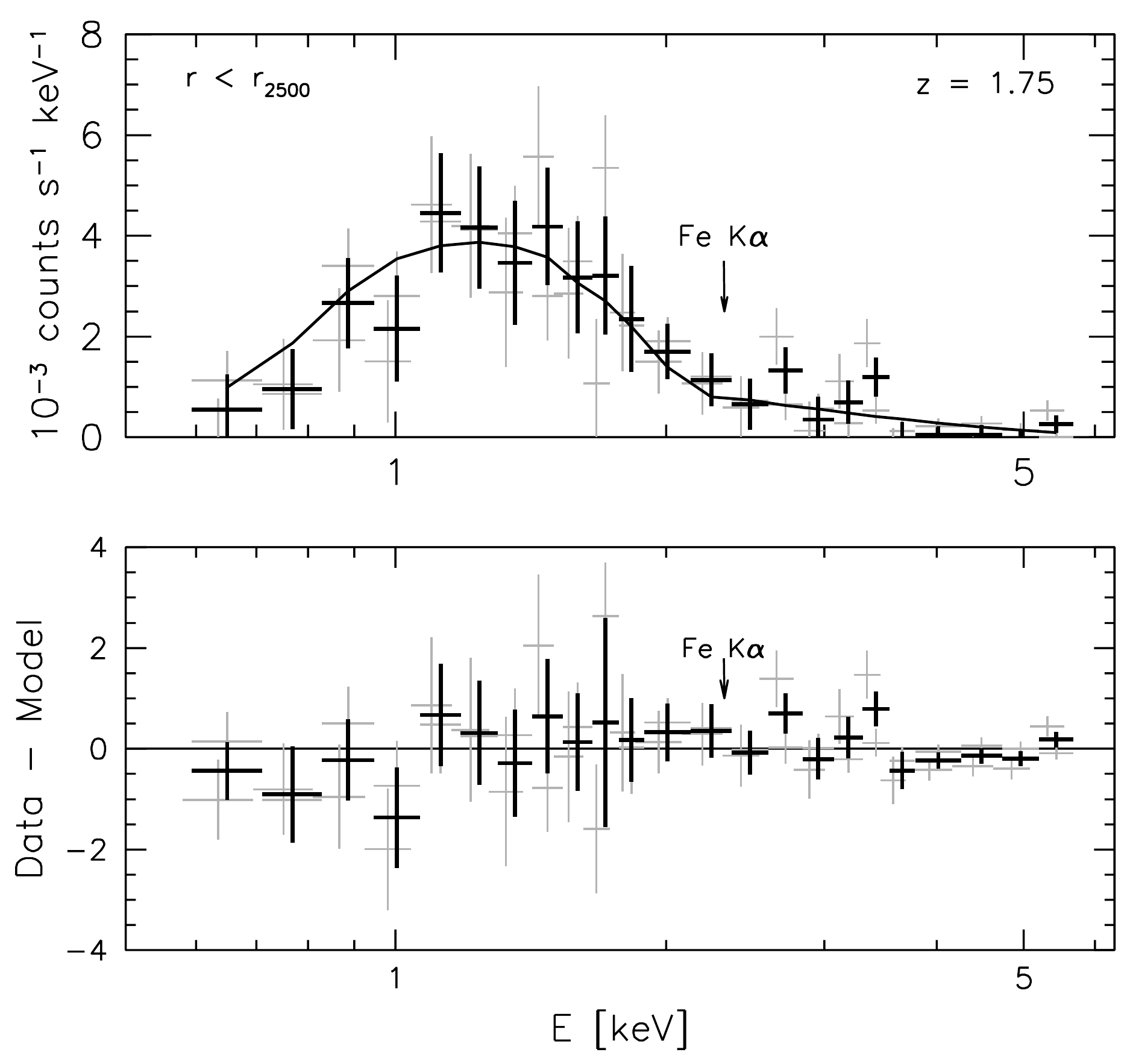}
\caption{{\it Upper panel}: Energy spectrum of \cl\ interior to
  $r_{2500}$.  The gray points are the two separate observations to
  which the models are fit, and the black points show the stacked
  spectrum.  The global temperature in this central region, based on
  the overplotted best-fit model, is $kT_{2500} = 6.2^{+1.9}_{-1.0}$
  keV.  The expected location of the Fe K$\alpha$ line is indicated,
  though we do not detect it.  The 1 $\sigma$ limit on the central
  metallicity is $Z_{2500} < 0.33 \,Z_\odot$, as dicussed in
  \textsection\ref{Sec: Z}.  {\it Lower panel}: Residuals of the
  best-fit model.}
\label{Fig: spectrum}
\end{figure}

The 0.5--6.0 keV ACIS-I spectra within $r_{2500}$
\citep[$\sim$0.45$r_{500,Y_X}$;][]{vikhlinin06} for each OBSID are
shown in Figure \ref{Fig: spectrum}.  This aperture was chosen to
maximize signal to noise, although we also consider a core-excised
(0.15--1)$r_{500}$ annulus below.  The spectroscopic temperature was
measured by modeling the X-ray spectrum using a combination of an
absorbed, optically thin plasma (\textsc{phabs x apec}), an absorbed
hard background component (\textsc{phabs x bremss}; kT = 40 keV), and
a soft, Galactic background component (\textsc{apec}; kT = 0.15
keV). Foreground and background models were constrained by fitting
simultaneously to an off-source region within the same field of view
and to the on-source region. The Galactic hydrogen column density,
N$_{\rm H}$, was set to the weighted average from the
Leiden-Argentine-Bonn survey \citep{kalberla05}. The source redshift
was fixed to $z=1.75$ \citep{stanford12}.  The reduced $\chi^2$ of
this fit is 0.95, demonstrating that the data are very well described
by the simple, single-temperature plasma model.  While the
goodness-of-fit is excellent, the individual parameters (e.g., $kT$,
$Z$) are not well constrained, as discussed below.
 
The average temperature within $r_{2500}$ is $kT_{2500} =
6.2^{+1.9}_{-1.0}$ keV. This global (not core-excised) temperature,
while not optimal as a mass observable due to inclusion of the cluster
core, is useful for comparison to lower resolution measurements from
other facilities or low SNR \chandra\ observations for which all the
counts are required to measure the temperature. For completeness we
also report a core temperature, within $r < 0.15\,r_{500}$, of
$kT_{\rm core} = 7.3_{-1.3}^{+2.7}$ keV.

We measured the core-excised spectroscopic temperature over $(0.15-1)$
\rfive\ using the T$_X$--M$_{500}$ scaling relation from
\citet{vikhlinin09} to estimate M$_{500}$, iteratively updating the
radius (\rfive) until it converged.  We used the pipeline described in
Benson et al.\ (in prep.) and \citet{mcdonald13}, which closely
follows the procedures described in \citet{andersson11}.  We find
$kT_{500} = 7.6_{-1.9}^{+8.7}$ keV, in good agreement with the global
value given above.  This temperature corresponds to a mass of
\Mfivetx\ $= 3.3_{-1.2}^{+5.7} \times 10^{14}$ M$_{\odot}$ and
\rfivetx\,$= 560 \pm 20$ kpc, where the subscript refers to the
scaling relation from which the physical quantity was derived.  The
error on the core-excised $kT_{500}$ is larger than that on
$kT_{2500}$ as roughly a third of the signal is removed with the
core, and the noise is significantly increased by moving from
$r_{2500}$ to $r_{500}$.  Nevertheless, temperature measurements
  and confidence level determinations remain unbiased down to count
  levels well below those in this work \citep[\eg][]{churazov96}.

\subsection{Gas Mass}

\begin{figure}[bthp]
\epsscale{1.15} 
\plotone{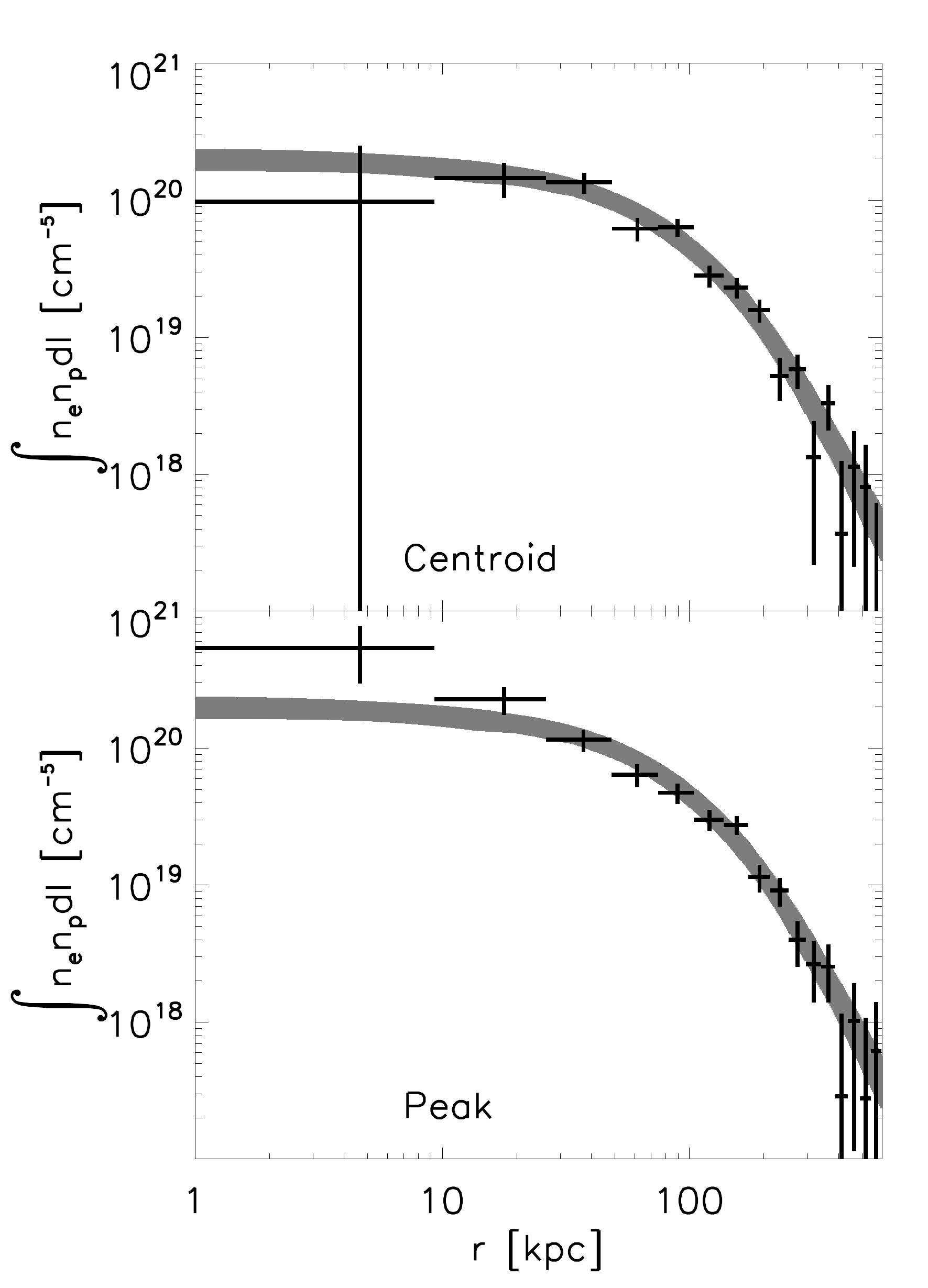}
\caption{{\it Upper panel}: X-ray surface brightness profile for \cl,
  derived using the large-scale X-ray centroid as the center.  The
  gray curve represents the best-fit projected double-beta model.
  {\it Lower panel}: X-ray surface brightness profile, centered on the
  X-ray peak.  The gray curve is the same as in the upper panel.
  There is evidence of a dense core on scales of $\sim20$ kpc, which
  appears slightly offset from the large-scale centroid (see
  Fig.~\ref{Fig: images}).}
\label{Fig: SBF}
\end{figure}

Similarly, M$_{g,500}$ was derived via the $f_g$--M$_{500}$ scaling
relation from \citet{vikhlinin09}. The derivation of the gas density
profile, described in detail in \citet{mcdonald13}, involves measuring
the X-ray surface brightness in the rest-frame energy 0.7--2.0 keV as
a function of radius.  The resulting surface brightness profile, shown
in Figure \ref{Fig: SBF} ({\it upper panel}), is fit with a projected
double-beta model, following \citet{vikhlinin06}. In converting from
electron density to gas density, we assume $\rho_g = m_p n_e A/Z$,
where $A$ = 1.397 is the average nuclear charge and $Z$ = 1.199 is the
average nuclear mass. We integrate the deprojected gas density profile
within an initial radius of \rfivetx, estimate the total enclosed gas
mass, and then use this to derive M$_{500}$ via the $f_g$--M$_{500}$
relation.  This process is iterated until convergence, leading to a
revised estimate of \rfivemg\ $= 500 \pm 10$ kpc and M$_{500,M_g} =
2.3_{-0.5}^{+0.7} \times 10^{14}$ M$_{\odot}$, consistent with
values inferred from the T$_X$--M scaling relation.  The gas mass is
M$_{g,500} = 2.5_{-0.6}^{+0.8} \times 10^{13}$ \msun.

\subsection{$Y_X$}
\label{Sec: Yx}

The product of the core-excised temperature and gas mass, referred to
as Y$_X$, approximates the total thermal energy in the cluster.  It
has been shown in simulations to be a low-scatter mass proxy that is
independent of the dynamical state of the cluster \citep{kravtsov06}.
Following the same approach as for $kT_{500}$ and M$_{g,500}$, we
estimate Y$_{X,500}$ iteratively, adjusting $r_{500}$ to satisfy the
Y$_X$--M$_{500}$ relation. We find $r_{500,Y_X} = 530 \pm 10$ kpc and
M$_{500,Y_X} = 2.6_{-0.5}^{+1.5} \times 10^{14}$
M$_{\odot}$. Within $r_{500,Y_X}$, we measure $Y_{500} =
1.9_{-0.6}^{+1.9} \times 10^{14}$ M$_{\odot}$ keV.

\subsection{Metallicity}
\label{Sec: Z}

Interior to $r_{2500}$ and $r_{500}$ (non core-excised), we measure
1$\sigma$ (2$\sigma$) upper limits on the metallicity of Z$_{2500}$
$<$ 0.33 Z$_{\odot}$ (0.47 Z$_{\odot}$) and Z$_{500}$ $<$ 0.10
Z$_{\odot}$ (0.18 Z$_{\odot}$), respectively.  As Figure \ref{Fig:
  spectrum} illustrates, there is no obvious detection of the Fe
K$\alpha$ emission line, such that, within $r_{2500}$, these data are
consistent with anywhere from ``typical'' to a complete absence of
metals.  Over larger radii we may be observing a marked lack of
metals, with the typical average metallicity of low-redshift clusters
being $Z_{500} \sim 0.3 \pm 0.1$ Z$_{\odot}$ \citep{degrandi01}, a
level of enrichment also seen in some $z>1$ clusters \citep{rosati09,
  santos12, degrandi14}.  However, deeper data are needed to properly
constrain the metallicity in this system, and determine whether it is,
indeed, metal-poor.\\

\section{Total Masses for \cl}
\label{Sec: Masses}

\subsection{M$_{500}$ from X-ray Scaling Relations}

Using the measured luminosity, temperature, gas mass and $Y_{500}$, we
reported in the previous section several complementary estimates of
the total mass within \rfive.  These masses, along with a host of
physical parameters, are listed in Table \ref{Tab: properties}.  Had
we assumed no evolution rather than a self-similar evolution, our
$Y_X$-based mass would be $\sim 50\%$ higher.

These masses all show a high level of consistency, indicating \cl\ is
relaxed, and offering good support for the hydrostatic approximation
that is underlying all these scaling relations.  A more stringent test
is provided by comparisons with complementary, independent mass
probes.

\begin{deluxetable}{lcc}
 \tablecaption{Properties of \cl \label{Tab: properties}}
  \tablewidth{0pt} \tablehead{ \colhead{Property} &
    \colhead{Value} & \colhead{Unit}  } \startdata
$z$                              & 1.75                &\\ 
X-Ray Centroid                   & (14:26:32.6, +35:08:25)                &  \\
X-Ray Peak Position              & (14:26:32.9, +35:08:26)             &\\
SZ Centroid                      & (14:26:34.0, +35:08:03)               &\\
BCG Position                     & (14:26:32.95, +35:08:23.6)               &\\\hline 
\rfivetx                       & $560 \pm 20$       & kpc\\ 
$r_{500,M_g}$                     & $500 \pm 10$       & kpc\\  
$r_{500,Y_X}$                     & $530 \pm 10$       & kpc\\  \hline
Counts                           & 420                   &\\ 
Count Rate                       &  4.14 $\times 10^{-3}$   & s$^{-1}$ \\ 
$S_{0.5-2}$                       & $(2.2 \pm 0.6) \times 10^{-14}$ & erg~s$^{-1}$~cm$^{-2}$\\ 
$L_{0.5-2}$                       & $(3.6 \pm 0.5) \times 10^{44}$ & erg~s$^{-1}$ \\ 
\Lxbol                       & $(12.8 \pm 1.1) \times 10^{44}$ & erg~s$^{-1}$\\ 
$kT_{\rm core}$\tablenotemark{1}   & $7.3^{+2.7}_{-1.3}$      & keV \\  
$kT_{2500}$                       & $6.2^{+1.9}_{-1.0}$      & keV \\ 
$kT_{500}$                        & $7.6_{-1.9}^{+8.7}$      & keV \\ 
M$_{g,500}$                       & $2.5_{-0.6}^{+0.8} \times 10^{13}$    & \msun \\ 
Y$_{500}$                         & $1.9^{+1.9}_{-0.6} \times 10^{14}$    &  \msun\ keV\\ 
$Z_{2500}$                        & $ < 0.33 \,(1 \sigma)$ &  $Z_\odot$\\ 
$Z_{500}$                         & $ < 0.10 \,(1 \sigma)$ & $Z_\odot$\\ \hline 
\Mfiveyx                       & $2.6_{-0.5}^{+1.5} \times 10^{14}$  &   \msun\\
\Mfivetx                       & $3.3_{-1.2}^{+5.7} \times 10^{14}$  &   \msun\\
\Mfivemg                       & $2.3_{-0.5}^{+0.7} \times 10^{14}$  &   \msun\\
\Mfivelxbol                       & $(2.8 \pm 0.3) \times 10^{14}$  &   \msun\\ 
\Mfivesz                       & $(2.6 \pm 0.7) \times 10^{14}$  &   \msun\\
\Mfivearc\tablenotemark{2}     & $(1.9_{-0.5}^{+0.7} - 2.6_{-0.7}^{+0.9}) \times 10^{14}$  &   \msun
\enddata
\tablenotetext{1}{The core temperature is measured within $r <
  0.15\,r_{500}$.}
\tablenotetext{2}{This mass ranges account for possible source
  redshifts beween $4.5 < z < 6$, as described in the text.}
\end{deluxetable}

\subsection{M$_{500}$ from Complementary Measurements}

\cl\ is unique in that it is the highest redshift cluster, by far, for
which independent mass measurements are available in the X-ray, SZ and
from strong gravitational lensing analyses.  This rare confluence is
only possible due to its extreme mass --- it is the most massive
cluster known at $z>1.5$ --- and the fortuitous (and surprising)
presence of a background galaxy for strong lensing \citep{gonzalez12}.

\citet{brodwin12} reported a strong ($>5 \sigma$) SZ detection at 31
GHz with CARMA.  The resulting mass measurement, \Mfivesz\ $ = (2.6
\pm 0.7) \times 10^{14}$ \msun, is in excellent agreement with all of
the X-ray mass measures presented above.  Relative to the $Y_{SZ}$-M
scaling relation from \citet{andersson11} that was used in
\citet{brodwin12}, the normalization is expected to increase by
$\sim$25\% in $Y_{SZ}$, corresponding to a $\sim$15\% increase in mass
(Benson \etal, in prep.).  The increase is primarily due to two
factors --- a statistical shift resulting from a much ($\sim$5x)
larger SZ and X-ray cluster sample, and updated mass normalization
from more recent weak-lensing observations \citep{hoekstra15}.  The
close agreement of the SZ mass for \cl\ with the X-ray mass measures
will not be affected by this modest change.

\citet{gonzalez12} reported the discovery of a giant gravitational
arc, visible in Figure \ref{Fig: images}, about 15\arcsec\ N of the
BCG at a clustercentric radius of $\sim$125 kpc.  Attempts to secure a
spectroscopic redshift for the source galaxy were unsuccessful, though
from the available photometry \citet{gonzalez12} constrained the
source redshift to the range $2 \la z_s \la 6$.  With a non-detection
in subsequent deep (AB $\sim 28$, 10 $\sigma$) \hst/ACS F606W imaging,
new data described in (Mo \etal\ in prep.), the redshift constraint is
now refined to $4.5 \la z \la 6$.  The strong lensing mass directly
measured within the arc radius ranges from $(6.9 \pm 0.3) \times
10^{13}$ \msun\ for $z_s = 6$ to $(8.5 \pm 0.3) \times 10^{13}$ \msun\
for $z_s = 4.5$.  Extrapolating to \rfive\ using the \citet{duffy08}
mass-concentration relation, this corresponds to a mass range between
\Mfivearc\ $= 1.9_{-0.5}^{+0.7} \times 10^{14}$ \msun\ and \Mfivearc\
$= 2.6_{-0.7}^{+0.9} \times 10^{14}$ \msun, in good agreement with all
the ICM-based masses.

A weak lensing analysis of \cl\ is underway (Mo \etal\ in prep.).  A
shear signal is detected in Cycle 20 \hst\ images, consistent with
expectations for a cluster this massive, even at $z=1.75$.

\begin{figure*}[bthp]
\epsscale{1.15} 
\plotone{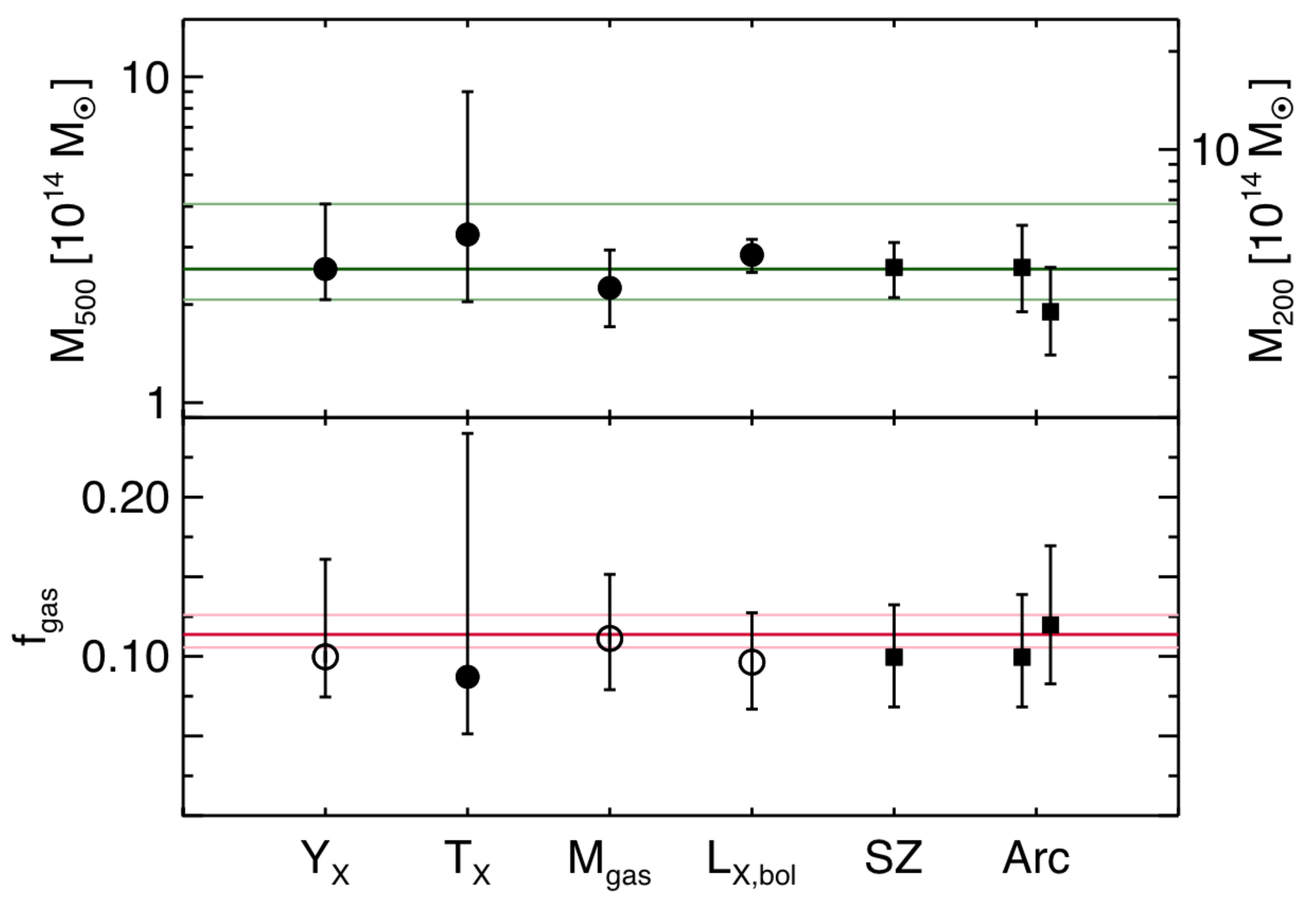}
\caption{{\it Upper panel}: Masses for \cl\ from the X-ray
  measurements described herein (filled circles), along with
  complementary SZ \citep{brodwin12} and strong lensing
  \citep{gonzalez12} masses (filled squares).  For the latter we plot
  the two masses corresponding to the extreme source redshift values,
  as described in the text.  We take the $Y_X$-based mass our best
  estimate (green horizontal line and error range), as it has the
  lowest intrinsic scatter \citep{kravtsov06}.  The remarkable
  agreement among these independent methods demonstrates that \cl\ is
  a remarkably mature, evolved cluster even at $z=1.75$.  {\it Lower
    panel}: \fgas\ measurements for \cl\ using each of these halo mass
  measurements.  \fgas\ measurements with independent M$_{g,500}$ and
  halo mass measurements have filled symbols; those affected by
  covariances between the gas and total mass are shown with open
  symbols.  The expected \fgas\ value for \cl, shown as the red line
  with 1 $\sigma$ errors, is taken from the \fgas-\Mfive\ relation of
  \citet{andersson11}.  The observed value of \fgas\ is in good
  agreement with the predicted value for all mass probes.}
\label{Fig: mass_fgas}
\end{figure*}
 
Figure \ref{Fig: mass_fgas} ({\it upper panel}) compares all of the
mass measures described in this paper, and shows the uniformly
excellent agreement among them.  This confirms that, despite its
extreme redshift, \cl\ is by all measures a relatively evolved,
relaxed cluster.  We take the low-scatter $Y_X$-based mass as our best
estimate of the halo mass of \cl, \Mfiveyx\ $ = 2.6_{-0.5}^{+1.5}
  \times 10^{14}$ \msun.

\subsection{$f_{\rm gas}$}

\noindent As massive galaxy clusters assemble, they are expected to
retain baryon fractions $(f_b)$ close to (but slightly lower than) the
universal value, with relatively low scatter
\citep[\eg][]{kravtsov12}.  Observationally, the gas mass fraction
$(f_{gas})$ is highest in the most massive clusters, with a weakly
decreasing fraction to lower masses \citep[\eg][]{vikhlinin09,
  andersson11}.
 
We calculate \fgas\ using each of the masses derived above,
integrating the gas mass out to the \rfive\ value appropriate to each
mass proxy.  These are plotted in Figure \ref{Fig: mass_fgas} ({\it
  lower panel}).  The \fgas\ measurements for which the M$_{g,500}$
and halo mass measurements are independent are shown as filled
symbols, while those affected by covariances between the gas and total
mass have open symbols.  We compare the measured \fgas\ values with
the value predicted from the redshift-independent \citet{andersson11}
relation.  This is shown as the solid red line, where the error range
includes the errors from \fgas--$M_{500}$ fit parameters in
\citet{andersson11}, as well as the mass error in \Mfiveyx, which we
took as the value of \Mfive\ for this relation.
 
The observed value of \fgas\ from every available mass proxy is
completely consistent with the expected value for a cluster with the
mass of \cl.  This provides additional evidence of the maturity of
\cl\, and bolsters our confidence in our use of scaling relations at
this extreme redshift to estimate its mass.  The recent measurement of
the gas fraction in XDCP J0044.0-2033 at $z=1.58$ \citep{tozzi15},
\fgas $= 0.08 +/- 0.02$, is slightly lower than mean of the present
measurements, though in agreement within 1 $\sigma$.

\section{Discussion}
\label{Sec: Discussion}

\subsection{Comparison With Other Distant, Massive Clusters}

\cl\ was the first relaxed, massive galaxy cluster to be confirmed at
$z>1.5$ \citep{stanford12}.  The present analysis of its X-ray
properties confirms the SZ and lensing masses reported in
\citet{brodwin12} and \citet{gonzalez12}, respectively.  The former
paper also demonstrated that \cl\ is an evolutionary precursor to the
most massive known clusters at all redshifts.  Several new $z>1.5$
clusters have subsequently been reported, but none are as massive, and
hence rare, as \cl.  Although the probability of detecting a cluster
this massive in the 9 deg$^2$ IDCS area is very small ($<1\%$),
\citet{brodwin12} demonstrated that its existence poses no threat to
$\Lambda$CDM. That paper also estimated that the SPT should expect
$\sim 2.4$ such clusters over their entire 2500 deg$^2$ survey.  In
the final SPT catalog paper, \citet{bleem15} reported three clusters
with strong decrements that do not yet have optical/IR confirmations.
Though not confirmed to date, current photometric limits suggest that
these three clusters lie at $z = 1.7 \pm 0.2$.

\citet{tozzi15} described deep \chandra\ observations of XDCP
J0044.0-2033 at $z=1.58$, from which they measured a total mass of
\Mfiveyx $=2.2_{-0.4}^{+0.5} \times 10^{14}$ \msun, using the same
\citet{vikhlinin09} Y$_X$--M scaling relation employed above.  This is
slightly below the identical mass measure for \cl, \Mfiveyx
$=2.6_{-0.5}^{+1.5} \times 10^{14}$ \msun, as well as all the other
ICM-based (i.e. X-ray and SZ) mass measures in Table \ref{Tab:
  properties}.

A direct comparison of temperatures is not straightforward, as
\citet{tozzi15} do not quote a value for $T_{2500}$ or $T_{500}$.
They instead measure a spectroscopic temperature of $kT =
6.7_{-0.9}^{+1.3}$ keV at a radius of 375 kpc.  As the \rfive\ values
for both clusters are quite similar, we measure the temperature at
this metric radius in \cl\ in order to make a meaningful comparison.
We find $kT_{375 \rm kpc} = 7.3_{-2.0}^{+3.4}$ keV, confirming that
\cl\ is likely the hotter and more massive cluster.

\citet{newman14} presented the spectroscopic confirmation of JKCS 041
at $z=1.80$, for which \citet{andreon14} report a mass, based on its
X-ray and optical properties, in the range \Mfive\ $\sim (1-2) \times
10^{14}$ \msun.  Given the non-detection in deep SZ imaging
\citep{culverhouse10}, the mass is likely at or below the lower end of
that range and thus less massive than \cl.

Finally, cluster candidates reported at $1.6 \la z \la 2$ by
\citet{papovich10}, \citet{zeimann12}, \citet{gobat11,gobat13} and
\citet{mantz14}, are all considerably less massive than \cl.
\citet{mantz14} describe the most massive of these, XLSSU
J021744.1-034536 with \Mfive\ $\sim 1-2 \times 10^{14}$ \msun, at a
photometric redshift of $z_{\rm phot} \sim 1.9$.  Despite this
redshift uncertainty, which strongly affects mass proxies in the
X-ray, we can say with certainty that this cluster has a lower mass
than \cl.  This assertion is based on a comparison of the spherically
averaged dimensionless Comptonization, $Y_{\rm sph, 500}$, measured
with CARMA for both clusters.  The SZ mass proxy, $M \propto (Y_{\rm
  sph, 500}\, D_A^2/E(z)^{2/3})^{3/5}$ \citep[\eg][]{marrone12}, where
$D_A$ is the angular diameter distance and $E(z)$ is evolution of the
Hubble parameter, is weakly dependent on redshift.  With $Y_{\rm sph,
  500}$ being a factor of 2.6 higher in \cl, we find its mass is
larger by a factor of $\sim$1.9.

\subsection{Dynamical and Cooling State of the ICM}
\label{Sec: CC}
 
In Figure \ref{Fig: SBF} we show the projected surface brightness
profile for two different choices of center: the large-scale
(250--500\,kpc) centroid of the X-ray emission and the X-ray
peak. Using the former definition, we fit a projected beta model to
the data, finding no evidence for a central surface brightness
excess. However, we show in the lower panel of Figure \ref{Fig: SBF}
that the peak of the X-ray emission --- which lies $\sim$30\,kpc from
the centroid --- represents a significant overdensity. Following
\cite{vikhlinin07}, we measure the cuspiness of the peak-centered
surface brightness profile, finding $\alpha \equiv (d\log\rho_g/d\log
r) \mid_{0.04r_{500}} = 0.82 \pm 0.09$. Such a high cuspiness at high
redshift is rare.  Indeed, \cite{vikhlinin07} found no such systems at
$z>0.5$ in their 400 deg$^2$ survey. Two similar systems, albeit at
much lower redshift ($z\sim1.1$), have been identified --- one by
\cite{mcdonald13} in a sample of 83 SPT clusters, and one by
\cite{santos12} in the WARPS survey.
 
The offset of $\sim$30 kpc between the X-ray centroid and the dense
core suggests that this cluster has undergone a recent interaction,
and that the cluster core is sloshing about the potential
minimum. Such an offset ought to remain visible for $\lesssim$\,500
Myr after any interaction \citep{ascasibar06,zuhone10}. Considering
that this system is being observed when the Universe was only 3.8 Gyr
old, and that it had to assemble rapidly to achieve such a high mass
at such early times, it is unsurprising that it retains an imprint of
this hurried growth in its core.

Following \cite{mcdonald13}, we calculate a pseudo-deprojected
entropy, using the deprojected gas density profile (\S3.5), an
aperture temperature, and assuming the X-ray peak as the center. We
find $K_0 \sim 20$ keV cm$^2$, corresponding to a cooling time of
$\sim$\,160 Myr. These properties are typical of cool core clusters
\citep{hudson10, mcdonald13}, suggesting that such systems can form
very early in the cluster lifetime.

The fact that the dense core appears elongated (Figure \ref{Fig:
  images}) suggests that, rather than a traditional cool core, we may
be observing a dense infalling group. This is consistent with many of
the observed qualities, including the low entropy, the offset from the
potential minimum and the non-symmetric morphology.  Conversely, the
nearly coincident BCG and X-ray peak (within $\sim 3\arcsec$; Figure
\ref{Fig: images}) suggests that this may be an offset core rather
than an infalling group.  Distinguishing between these two scenarios
requires a deeper X-ray follow-up observation.

\section{Conclusions}
\label{Sec: Conclusions}

We have presented a deep 100 ks \chandra\ observation of \cl\ at
$z=1.75$, the most massive cluster discovered at $z>1.5$ from {\it
  any} method.  We measured the luminosity, temperature and gas mass,
from which we derived halo mass estimates from the T$_X$--M, $f_g$--M,
Y$_X$--M, and L$_X$--M scaling relations.  These all show excellent
consistency and are in remarkable agreement with independent SZ and
strong lensing masses.  Similarly, the gas mass fractions for all
these mass proxies were found to be in good agreement with each other
and with the value predicted from low redshift clusters.

The bolometric luminosity is \Lxbol\ $= (12.8 \pm 1.1) \times 10^{44}$
erg~s$^{-1}$, from which we estimate a mass of \Mfivelxbol\ $ = (2.8
\pm 0.3) \times 10^{14}$ \msun.  We measure a central temperature
within $r_{2500}$ of $kT_{\rm 2500} = 6.2^{+1.9}_{-1.0}$ keV and a
core-excised temperature within \rfivetx\ $= 560 \pm 20$ kpc, used in
mass scaling relations, of $kT_{500} = 7.6_{-1.9}^{+8.7}$ keV.  This
results in a mass of \Mfivetx\ $ = 3.3_{-1.2}^{+5.7} \times 10^{14}$
\msun.  We find no evidence for metals in the ICM of this system,
placing a 2$\sigma$ upper limit of $Z_{500} < 0.18\,Z_{\odot}$,
suggesting that this system may still be in the process of enriching
its intracluster medium.

We measure a gas mass of M$_{g,500} = 2.5_{-0.6}^{+0.8} \times
10^{13}$ \msun, from which we infer an $M_g$-based halo mass of
\Mfivemg $ = 2.3_{-0.5}^{+0.7} \times 10^{14}$ \msun.  From the gas
mass and core-excised temperature, we find Y$_{500} =
1.9^{+1.9}_{-0.6} \times 10^{14}$ \msun\ keV, from which we derive our
lowest-scatter estimate of the mass of \cl, \Mfiveyx $ =
2.6_{-0.5}^{+1.5} \times 10^{14}$ \msun.  All the X-ray masses for
\cl\ are in agreement, as are the SZ- and lensing-based mass from
previous analyses.
 
The cluster has a dense, low-entropy core, offset by $\sim$30 kpc from
the X-ray centroid, which makes it one of the few ``cool core''
clusters discovered at $z>1$, and the first known cool core cluster at
$z>1.2$. The offset of this core from the large-scale centroid
suggests that this cluster has had a relatively recent
($\lesssim$\,500 Myr) merger/interaction with another massive system.
We measure a central entropy of $K_0 \sim 20$ keV cm$^2$, indicating
the cool core may have a very rapid cooling time of $\sim$\,160 Myr.

\cl\ is the first cluster at $z>1.5$ to have all these independent
mass measurements. In addition to being the most massive known cluster
in this redshift regime, \cl\ also has a remarkably relaxed ICM
suggesting a very early and rapid formation.  Deeper follow-up X-ray
observations are essential to permit a meaningful constraint on its
metallicity and to measure the entropy profile to better understand
how and when such cores can form and when the cooling/feedback loop is
established.

\acknowledgments 

Support for this work was provided by the National Aeronautics and
Space Administration (NASA) through \chandra\ Award Number GO3-14135A
issued the the {\it Chandra X-ray Observatory} Center, which is
operated by the Smithsonian Astrophysical Observatory for and behalf
of NASA under contract NAS8-03060.  This work is based in part on
observations made with the {\it Spitzer Space Telescope}, which is
operated by the Jet Propulsion Laboratory, California Institute of
Technology under a contract with NASA.  Support for HST programs
11663, 12203 and 12994 were provided by NASA through a grant from the
Space Telescope Science Institute, which is operated by the
Association of Universities for Research in Astronomy, Inc., under
NASA contract NAS 5-26555.

We thank Alexey Vikhlinin, Bradford Benson, Daniel Marrone and Adam
Mantz for helpful discussions.  We are grateful to the referee for a
careful reading that improved the paper.  This work would not have
been possible without the efforts of the support staffs of the {\it
  Chandra X-ray Observatory}, and the {\it Spitzer} and {\it Hubble
  Space Telescopes}.

{\it Facilities:} \facility{Chandra (ACIS-I)}, \facility{HST (ACS,
  WFC3)}, \facility{Spitzer (IRAC)} 

\bibliographystyle{astron2} \bibliography{bibfile}

\end{document}